\documentclass[11pt,twoside]{article}

\usepackage{asp2006}
\usepackage{epsf}

\begin{document}
\title{Stellar jets} 
\author{Thomas J. Maccarone}
\affil{School of Physics and Astronomy, University of Southampton, Highfield, Southampton, SO16 4ES, United Kingdom} 

\begin{abstract} 
With a goal of understanding the conditions under which jets might be
produced in novae and related objects, I consider the conditions under
which jets are produced from other classes of accreting compact
objects.  I give an overview of accretion disk spectral states,
including a discussion of in which states these jets are seen.  I
highlight the differences between neutron stars and black holes, which
may help give us insights about when and how the presence of a solid
surface may help or inhibit jet production.

\end{abstract}


\section{Introduction}   

Jets have been seen from a variety of classes of astrophysical
objects.  The first report of an astrophysical jet was a remark that a
streak of light seemed to be emanating from the galaxy M87 (Curtis
1918), but studies of jets did not begin in earnest until the
development of radio astronomy.  It has long been appreciated that
active galactic nuclei often power relativistic jets.  In the past
decade or so, a picture has been emerged in which it seems that a
rotating magnetic field is sufficient to power a jet -- jets have been
seen from all sorts of accreting stars, like young stellar objects,
supersoft sources, accreting stellar mass black holes, and accreting
neutron stars in low mass X-ray binaries.  Jets have also been seen
from a few young rotation-powered pulsars, indicating that an
accretion disk is not a necessary element for jet production.  The
only classes of objects from which jets have not been widely reported
are accretion powered pulsars in high mass X-ray binaries (where the
magnetic field truncates the accretion disk far from the surface of
the accreting neutron star), and cataclysmic variables (where a few
shallow non-detections of radio emission in the 1970's led to a
general disinterest in making deeper searches).  In this paper, I will
review the literature of jets from accreting stellar mass black holes
and accreting low magnetic field neutron stars.  I will then use these
results to develop some intuition about when and why jets might be
seen from accreting white dwarfs.  For a more general review of
astrophysical jets, I suggest Maccarone \& K\"ording (2006) for
readers looking for a gentle introduction to the topic, Livio (1999)
for readers looking for a greater level of technical detail, and
Fender (2006) for readers looking for a more detailed view of jets
from stellar mass compact objects.

\section{Spectral states of disks}

For the purposes of establishing a connection between accretion and
ejection, stellar mass black holes are the most useful objects.  They
vary in luminosity by factors of $\sim10^7$ on timescales of years.
As their emission comes out primarily in X-rays, they can be observed
with all-sky monitors, so outbursts from new sources can be detected
easily across most of the Galaxy.  

In order to understand better the conditions in an accretion flow
which allow jet production, it is necessary to understand some of the
basics of accretion disk phenomenology.  Black hole X-ray binaries
exhibit a spectral state phenomenology, much of which has clear
analogies to the spectral state phenomenology of neutron star X-ray
binaries.  These states are defined by qualitative patterns of
behavior of both the spectral energy distribution and the Fourier
power spectrum of the sources' X-ray emission.

The simplest of the spectral states in accreting black holes is the
high/soft state.  In this state, a black hole's X-ray spectral energy
distribution is well described by a multi-temperature disk model --
essentially the disk solution first proposed by Shakura \& Sunyaev
(1973) and Novikov \& Thorne (1973).  These disks appear to extend to
the innermost stable circular orbits of the accreting black holes
(e.g. Sobczak et al. 1999).  In most cases a weak power law tail to
the spectrum is detected, sometimes extending out to several MeV
(McConnell et al. 2000), but never is it energetically important.

The next simplest state is the low/hard state, in which most accreting
black holes spend most of the time.  In this state, the X-ray spectral
energy distribution is reasonably well modelled by an exponentially
cutoff power law, with a photon index of 1.5-1.9, and the cutoff at
about 100-200 keV.  This type of spectrum is easily produced by
thermal Comptonization (Thorne \& Price 1975; Sunyaev \& Tr\"umper
1979) by a cloud of hot ($k_B$${T}\approx70$keV) with an optical depth
close to unity.

A third, hybrid state is seen at state transitions, and sometimes
during extended period during which sources remain at luminosities
very close to the Eddington limit.  This state is called the very high
state when it is seen at high luminosities, and the intermediate state
when it is seen at lower luminosities.  The X-ray spectra of these
states show strong quasi-thermal components and strong power law
components.  The power law componets in these states are steep
(i.e. $\Gamma=2.5-3.0$), like the power law components in soft states.
Generally, however, the properties of these two states, both in terms
of X-ray spectra and variability, there are no clear differences apart
from luminosity between what is generally called the very high state
and what is generally called the intermediate state (see e.g. Homan et
al. 2001).

\subsection{Timing states of disks}

Since the days of EXOSAT, when it became possible to make good power
spectra of a large number of accreting sources at a variety of
accretion rates, it has been clear that the timing properties of
accreting black holes and neutron stars fit a set of patterns
(e.g. van der Klis 2005 and references within).  More recently, this
picture has been extended to include, at least at some level, white
dwarfs as well (Mauche 2002; Warner, Woudt \& Pretorius 2003),
indicating again that much of the physics of accretion is generic to
disks, rather than being peculiar to objects where relativistic
effects are important.  There are some power spectral features which
clearly are associated with only one class of source, such as the high
frequency quasi-periodic oscillations seen with 2:3 frequency ratios
in some black hole accretors and are seen at frequencies high enough
to require relativistic effects (e.g. Strohmayer 2001; Abramowicz \&
Kluzniak 2001), but these are generally easily isolated from the
global trends in the power spectra.

Low/hard state sources characteristically have higher amplitudes of
variability than high/soft state sources, but lower characteristic
frequencies of variability (e.g. van der Klis 1994).  During the state
transitions, and in the persistent very high states, strong
quasi-periodic oscillations are often seen (van der Klis 1994).  The
geometrically thin accretion disks are stable against short timescale
variability -- this has been determined observationally through
multiple techniques (Churazov, Gilfanov \& Renivtsev 2001; Maccarone
\& Coppi 2002).  The small amplitude of variability in the high/soft
state is driven by the component which produces the weak hard tail in
the X-ray spectrum.  Given this result, the relative variability
properties of the hard and soft states agree with the picture --
variability is produced in the hard X-ray emitting region, and the
characteristic timescale of the variability is related to some
chracteritic spatial scale in the emission region.  The strong
quasi-periodic oscillations in the state transitions are less well
understood.

\subsection{When do the state transitions occur?}

Transitions between spectral states do not appear at a fixed fraction
of the Eddington luminosity as was predicted by the earliest
theoretical work.  Instead, a hysteresis behavior has been found from
most X-ray binaries in their state transition luminosities
(e.g. Miyamoto et al. 1995; Maccarone \& Coppi 2003).  Typically,
state transitions from the hard state to the soft state occur at
luminosities about five times higher than the luminosities on the
return from the soft state to the hard state.  Transitions from the
soft state to the hard state happen typically at about 2\% of the
Eddington limit, with much less scatter than the transitions from the
hard state to the soft state (Maccarone 2003), but there are cases of
soft-to-hard state transitions at different luminosities in the same
source (Xue, Wu \& Cui 2006; Yu \& Dolence 2006).

A particularly interesting recent result which has received remarkably
little attention is that white dwarfs show spectral state transition
phenomenology which is quite similar to that seen from accreting black
holes and neutron stars -- both harder emission at low luminosities
than high luminosities (which had been known for some time), but also
similar hysteresis loops in hardness versus intensity (McGowan,
Priedhorsky, Trudolyubov 2004; Wheatley, Mauche \& Mattei 2003). The
models invoked for the hard spectral states in white dwarfs are
generally not the same as those invoked in black hole and neutron star
accretors (although possible similarities have been appreciated by
some authors for about a decade -- see for example Meyer \&
Meyer-Hofmeister 1994; Meyer-Hofmeister \& Meyer 1999 which invoke
essentially the same physics to describe the outburst properties and
state transitions of dwarf novae and of soft X-ray transients with
black hole accretors), but the similarities in the state transition
properties, combined with the timing similarities between dwarf novae
and X-ray binaries together imply that much of what is seen is
universal to accretion disks, rather than peculiar to accretion disks
with a particular type of compact object.

\section{Jets from black hole X-ray binaries}

Three classes of jets have been seen from X-ray binaries with likely
black hole accretors.  The first class consists of the spatially
extended, highly variable but quasi-persistent jet emission from the
systems which have been studied the longest -- SS 433 and Cygnus X-3.
Both these systems suffer from high obscurration of the X-ray emission
from the source, likely due to a strong stellar wind in Cygnus X-3 and
to an edge-on geometrically thick accretion disk in SS 433.
Furthermore, both these source may be accreting more material than is
needed for them to reach the Eddington limit, and neither has a
reliable mass estimate, leaving open the possibility that one or both
actually has a neutron star accretor.  Thus while these two objects
are quite interesting, they seem unlikely to be the keys for
understanding the behavior of the broader class of X-ray binaries.

The two other classes of jets are seen from more normal soft X-ray
transients, as well as from the canonical black hole X-ray binary
Cygnus X-1.  These are steady jets in the low/hard state and the more
rapid jet ejections seen at state transitions.  It should be noted
that these rapid ejections are not seen in every outburst of every
soft X-ray transient.  Some outbursts do not reach a high enough
luminosity to undergo a state transition (e.g. Brocksopp,
Bandyopadhyay \& Fender 2004), with the typical peak outburst
luminosity in X-ray binaries, like in CVs, related roughly
monotonically to the orbital period of the system (Warner 1987;
Shahbaz,Charles \& King 1998; Portegies Zwart, Dewi \& Maccarone
2004).

Radio emission in low/hard state X-ray binaries is ubiquitous.  These
objects have flat-to-slightly inverted radio spectra
(i.e. $\alpha=0.0-0.3$, where $f_\nu \propto \nu^{\alpha}$, with
$f_\nu$ the flux density, and $\nu$ the frequency) typical of compact
conical jets (Blandford \& K\"onigl 1979; Hjellming \& Johnston 1988),
with a break typically in the infrared or optical (Russell et al. 2006
and references within).  It is worth noting that flat radio spectra do
not generically imply jets (such radio spectra are also seen, for
example, from HII regions and planetary nebulae, indicating that flat
spectrum sources need be neither due to synchrotron emission nor from
collimated regions), so additional evidence is required to prove that
this radio emission is from synchrotron emission in compact conical,
relativistic jets.

In Cygnus X-1 such jets have actually been imaged (Stirling et
al. 2001), and the data quality has generally not been good enough to
do so for other low/hard state black holes.  The brightness
temperatures implied by the fluxes of radio emission implied require
that the emission comes from a region much larger than the orbital
separation of the binary.  Furthermore, where good data exist, the
radio emission is seen to show weak, but significant linear
polarization with a steady polarization angle (Corbel et al. 2000),
also indicative of a jet.  The best (or at least most interesting, in
this writer's opinion) evidence that these jets are at least mildly
relativistic comes from time series analysis of XTE J1118+480, where
the lag between the X-ray emission and the optical emission gives a
rough estimate of the time it takes for changes in the accretion disk
where the X-rays are produced to propagate up the jet to the region
where the optical emission is produced (e.g. Malzac, Merloni \& Fabian
2003).

In low/hard states, the radio luminosity, $L_R$ from the jet scales
with the X-ray luminosity from the accretion flow $L_X$ such that
$L_R$$\propto$$L_X^{0.7}$ (Corbel et al. 2000; Gallo, Fender \& Pooley
2003).  Considerable scatter is seen about this relation.  Much of the
scatter likely derives from a variety of sources which may not reveal
much about the physics of jet production -- non-simultaneity of the
X-ray and radio data, uncertainties in the distances to sources, and
uncertainties in source masses.  However, there is clearly some
scatter due to reasons which are based in physics, and not in
observational difficulties.  Recently, it has been suggested that GX
339-4 shows parallel tracks in radio versus X-ray emission (Nowak et
al. 2005), and shown clearly that parallel tracks exist in the
infrared versus X-ray emission in XTE J1550-564 (Russell et al. 2007).
Additional parallel tracks were reported during ``failed state
transitions'' in Cygnus X-1 (Nowak et al. 2005).

In the high/soft states of black holes, radio emission has been
detected only once, and this appears to have been decaying emission
from a transient event that had taken place shortly before the
observation (Corbel et al. 2004).  The best searches for radio
emission from soft state black holes indicate that the jet power is
suppressed by a factor of at least 30-50 in the high/soft state
(Corbel et al. 2001).  In fact, that radio emission is well correlated
with hard X-rays and anti-correlated with soft X-rays was first
discovered about 35 years ago (Tananbaum et al. 1972), but has only
become well-appreciated in the last decade or so (Harmon et al 1995;
Fender et al. 1999).  In the early 1970's, this result was considered
exciting because the association of a state transition in the X-rays
with a dramatic change in the radio properties of a source in the
X-ray error box allowed the use of a radio position for searches for
the optical counterpart of Cygnus X-1.  Attempts to establish a
dynamically confirmed black hole were, not surprisingly, considered
more important than attempts to understand the radio emission from
stellar mass black holes.  It took the confluence of the discovery of
the ``microquasar phenomenon'' generating new interest in X-ray
binaries, along with the launches of CGRO and RXTE, which could
provide intensive flux monitoring of X-ray binaries, to go beyond
Tananbaum's original discovery to promote an understanding of jet
production based on X-ray/radio connections in X-ray binaries.

At the transitions from hard to soft states, there have very often
been high luminosity, high velocity jet ejections observed
(e.g. Mirabel \& Rodriguez 1994; Hjellming \& Rupen 1995).  An
intriguing possbility is that these events occur because the jet
velocity is tied to the escape speed at the inner ``edge'' of the
accretion disk, and, as the state transitions proceed, the inner edge
of the accretion disk moves inwards, raising the velocity of the jet,
and leading to external shocks of the newer high velocity jet material
against the old slow jet material ejecting during the long low/hard
states (Vadawale et al. 2003; Fender, Belloni \& Gallo 2003).

Some recent work (e.g. Rykoff et al. 2007 and references within) has
been suggested to indicate that the inner edge of the accretion disk
stays fixed at the innermost stable circular orbit even across state
transitions into the low hard state.  It is more likely that there are
not sudden and dramatic changes in the inner disk radius at the time
of the state transition, but rather smooth changes in the inner disk
radius as a function of luminosity throughout the low/hard state, with
the brightest low/hard states having inner disk radii quite close to
the innermost stable circular orbits.  Gradual transitions would
better explain the timing data -- the characteristic timescales change
smoothly throughout the low/hard state, especially in the rising
low/hard states (e.g. Pottschmidt et al. 2003; K\"ording et al. 2007)
-- and would even better explain the spectral data presented in Rykoff
et al. (2007): the best fitting inner disk radii are larger for the
low/hard states where there are sufficient counts to make constraining
spectral fits than they are in the high/soft state observations.  This
picture would also allow for the clear evidence of large inner disk
radii for quiescent black hole X-ray binaries without requiring the
disk to begin receding at some arbitrary luminosity below the state
transition luminosity.

\section{Thick disks and the theory of jet production}

The phenomenological association of radio emission with hard X-ray
emission is likely associated with a more direct, physically motivated
connection between geometrically thick accretion flows and jet
production.  It has been suggested that large scale height magnetic
fields are necessary for extracting rotational energy from accretion
disks and/or extracting the spin energy from a black hole (Livio,
Ogilvie \& Pringle 1999; Meier 2001), and these are the two most
prominent means for providing the energy needed to power jets with
accretion disks (e.g. Blandford \& Znajek 1977; Blandford \& Payne
1982).

\section{Jets from neutron star X-ray binaries}

Until very recently, the radio emission from neutron star X-ray
binaries has been considerably less well studied than that from black
hole X-ray binaries.  A number of factors have contributed to this
discrepancy.  The most important is that the relative faintness of the
neutron stars as radio sources, coupled with the locations of most of
the brightest neutron stars in the Southern Hemisphere where they are
unaccessible or barely accessible to the Very Large Array, has made
such studies technically challenging.  Despite these techincal
challenges, some major advances have been made in the past few years
in understanding the radio emission from neutron stars.

Radio observations of accreting neutron stars yield both key
similarities and differences.  For the low/hard state neutron stars
there is still a clear correlation between radio and X-ray luminosity,
as seen in the low/hard state black holes, but with a much stronger
dependence of radio power on X-ray power.  The best fitting power law
index for the radio/X-ray relation in accreting neutron stars is $L_R
\propto L_X^{1.4}$ (Migliari \& Fender 2006), twice as steep as the
relation for black holes.  This has been interpreted as evidence that
the black hole systems advect energy across their event horizons,
while the neutron star systems release this energy in their boundary
layers; the difference is naturally and exactly accounted for by a
radiative efficiency which is constant for neutron star accretors, and
which scales with $\dot{m}$ for black hole accretors, as predicted for
example, in advection dominated accretion flow models (e.g. Narayan \&
Yi 1994).

At higher luminosities, a key difference between neutron stars and
black holes presents itself.  In their equivalents of the high soft
state, neutron stars are detectable radio emitters.  Their radio
fluxes are well below the extrapolation of what might be expected
based on their X-ray fluxes and the relation in the hard state, so jet
production is, in some sense, suppressed in neutron stars in soft
states, but it is not suppressed by nearly as much as the black hole
systems are.  Figure 1 presents a schematic view of the similarities
and differences between jet properties in neutron stars and black
holes.

While much less is known about jets from white dwarf systems and their
connections with the properties of the underlying accretion flow,
there is some reason to believe that that qualitative behavior is
analogous to that from neutron stars.  The symbiotic star CH Cyg
showed a correlation between a jet ejection event and a dramatic
reduction in the amount of high frequency flickering (Sokoloski \&
Kenyon 2003), suggesting some agreement with this picture.  

\begin{figure}
\epsfysize=3.0in
\epsfbox{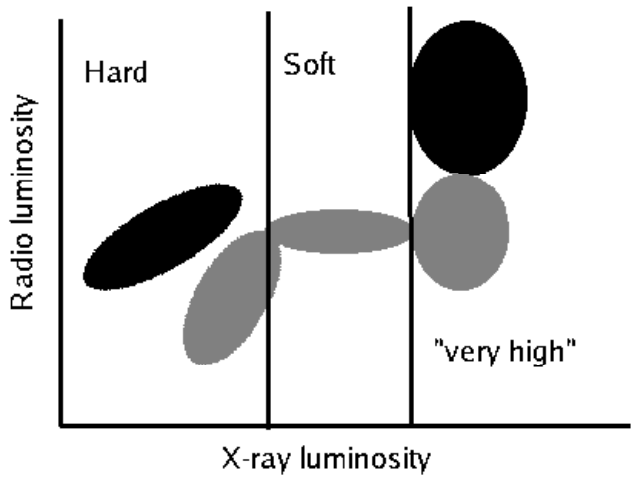}
\caption{A schematic diagram showing the similarities and differences
between neutron stars (grey regions) and black holes (black regions)
in their jet production properties.  The similarities are increasing
$L_R$ with increasing $L_X$ in low luminosity states, a deficit
compared to the extrapolation of the low/hard state trend (the left
hand portion of the diagram) in the high/soft states (enclosed within
the vertical lines), and a higher, but more scattered radio luminosity
in the ``very high'' states.  The most important difference is that
the soft state radio emission is not nearly as strongly suppressed in
neutron star systems as it is in black hole systems.}

\end{figure}

\section{Are boundary layers important for jet production?}

The radio emission in low luminosity neutron stars and black holes are
largely the same, with the differences being accounted for by the fact
that neutron stars cannot advect energy across their nonexistent event
horizons.  In these states, then, it seems most likely that a
geometrically thick accretion disk is providing the bulk of the energy
that goes into the jet.  In the higher luminosity states, where the
accretion disk is geometrically thin, the downturn in radio luminosity
is much stronger in the black hole systems than in the neutron star
systems.

It is most natural, then, to associate the jets seen from soft state
neutron stars with some means of launching jets that makes use of the
solid surface of the neutron star.  We had previously suggested that
this might be due to the magnetic field of the neutron star providing
a ``seed'' field (Maccarone \& K\"ording 2006).  However, what is
needed is a magnetic field with a scale height of order the radius of
the neutron star, and the $1/r^3$ dependence of dipole fields may
present a problem here, as might magnetic screening (Cumming, Zweibel
\& Bildsten 2001) which can dramatically reduce the effective magnetic
fields of high accretion rate neutron stars.

More likely then, is that the boundary layers of the surface of the
neutron star, which have strong differential rotation (a key
ingredient for dynamos) and have large scale heights (typically of
order the neutron star radius in the neutron star case -- Popham \&
Sunyaev 2001), can generate magnetic fields which can be used to power
jets.  Indeed, this suggestion has been raised in the past, as a
candidate for being the extra ``energy source'' suggested to be
necessary for jet production, apart from normal disk accretion (Livio
1999).  By its very nature, a boundary layer is a region with
differential rotation, implying that it would be reasonable for
dynamos to work effectively in a boundary layer.  When the accretion
flow is geomtrically thin, the amount of rotational energy which can
be extracted from the boundary layer exceeds the amount which can be
extracted from the thin disk.

This scenario, where the boundary layer provides the seed of the
magnetic field, thus implies that white dwarf accretion disks with
high accretion rates and relatively thick boundary layers should be
capable of powering jets; similarly, even supersoft sources and
symbiotic stars, where there need not be an accretion disk, should be
able to power jets though extrating some of their boundary layers'
rotational energy.

\section{Conclusions}

The jet properties of compact objects are set primarily by the class
of compact object, the state of the accretion disk, and the mass
accretion rate -- a good estimate of the jet spectrum and luminosity
can be made based on these parameters.  At the same time, the observed
hysteresis effects indicate that clearly some other information is
important for understanding the jet-disk connection.  Spectral state
phenomenology seems remarkably similar in accretion disks around black
holes and neutron stars, and perhaps even in disks around white
dwarfs.  While geometrically thick accretion disks seem to be the most
efficient way to power radio jets, the detection of reasonably strong
radio emission from some soft state neutron star systems indicates
that thick disks are not the {\it only} way to power such jets.  The
boundary layer seems the most likely source of jets in such systems,
and since powerful accretion onto boundary layers in white dwarf
systems is also seen, this is a possible explanation for jets seen
from high accretion rate white dwarfs.


\acknowledgements I am especially grateful to Elmar K\"ording for
discussions which went into our 2006 Astronomy \& Geophysics article,
which helped form the basis for many of the thoughts presented here,
as well as for numerous other useful discussions.  I am also grateful
for discussions with numerous colleagues and collaborators over the
years -- especially Rob Fender, Elena Gallo, Simone Migliari, Mike
Nowak and Dave Russell, and for enlightening discussions during this
workshop, especially with Michael Rupen and Jeno Sokoloski.


\end{document}